\documentclass[prl,twocolumn,superscriptaddress,longbibliography]{revtex4-1}
\usepackage{amssymb,amsmath,amsfonts,bm}
\usepackage{graphicx}
\usepackage{epstopdf}
\usepackage{times}
\usepackage{float}
\usepackage{lipsum}
\usepackage{color}

\newcommand*{\red}{\textcolor{red}} 

\usepackage{hyperref}
\hypersetup{colorlinks=true, linkcolor=blue, citecolor=red, urlcolor=blue  }

\usepackage{physics}

\begin{document}

\title{Super-Robust Non-Adiabatic Geometric Quantum Control}


\author{Bao-Jie Liu}
\affiliation{Department of Physics, Southern University of Science and Technology, Shenzhen 518055, China}

\author{Yuan-Sheng Wang}
\affiliation{Department of Physics, Southern University of Science and Technology, Shenzhen 518055, China}
\affiliation{School of Physical Sciences, University of Science and Technology of China, Hefei 230026, China}

\author{Man-Hong Yung}  \email{yung@sustech.edu.cn}
\affiliation{Department of Physics, Southern University of Science and Technology, Shenzhen 518055, China}
\affiliation{Shenzhen Institute for Quantum Science and Engineering, Southern University of Science and Technology, Shenzhen 518055, China}
\affiliation{Guangdong Provincial Key Laboratory of Quantum Science and Engineering, Southern University of Science and Technology, Shenzhen 518055, China}
\affiliation{Shenzhen Key Laboratory of Quantum Science and Engineering, Southern University of Science and Technology, Shenzhen,518055, China}

\date{\today}

\begin{abstract}
Non-adiabatic geometric quantum computation (NGQC) and non-adiabatic holonomic quantum computation (NHQC) have been proposed to reduce the run time of geometric quantum gates. 
However, in terms of robustness against experimental control errors, the existing NGQC and NHQC scenarios have no advantage over standard dynamical gates in most cases. 
Here, we give the reasons of why non-adiabatic geometric gates are sensitive to the control errors, 
and further, we propose a scheme of super-robust non-adiabatic geometric quantum control, 
in which the super robust condition can guarantee both high speed and robustness of geometric gate. 
To illustrate the working mechanism of super-robust geometric quantum gates, we give two simple examples of SR-NGQC and SR-NHQC for two- and three-level quantum systems, respectively. 
Theoretical and numerical results with the experimental parameters indicate that our scheme can  
significantly improve the gate performance comparing with the previous NGQC, NHQC and standard dynamical schemes.
Super robust geometric quantum computation can be applied to various physical platforms such as superconducting qubits, quantum dots, and trapped ions. 
All of these sufficiently show that our scheme provides a promising way towards robust geometric quantum computation.

\end{abstract}

\maketitle

\emph{Introduction}.\textbf{---} Realizing high-fidelity and fault-tolerant quantum gates is very essential for quantum information processing, since control errors and environment-induced noises are ubiquitous in operating real quantum devices.
Geometric quantum computation utilizes a unique property that the time-dependent quantum state would accumulate Abelian geometric phase~\cite{b2,b1}, or non-Abelian holonomy~\cite{Zanardi1999,gqc,b4} under a cyclic quantum evolution.  
The geometric phase and holonomy depend only on the global properties of the evolution trajectories. Consequently, geometric quantum gates are robust against local disturbances during the evolution~\cite{Zhu2005,Berger2013,Chiara,Leek,Filipp}. More specifically, geometric quantum computation can be divided into Abelian GQC and holonomic quantum computation (HQC) depending on whether the geometric phase is a real number~\cite{b2} or a matrix~\cite{b4} (non-Abelian holonomy).

Early applications of GQC are dependent on adiabatic quantum evolutions to suppress transitions between different instantaneous eigenstates of Hamiltonian \cite{Jones,Wu2013PRA,Huang2019,Duan2001a,lian2005}. 
Adiabatic GQC has been experimentally verified as a noise-resilient scenario against fluctuations of control parameters~\cite{Wu2013PRA,Huang2019}. However, adiabatic quantum dynamics implies lengthy gate time and thus long exposure time to the environment-induced decoherence. 
To overcome such a problem, non-adiabatic geometric quantum computation (NGQC)~\cite{b5,b6,Thomas,zhao2017,Li2020PRR,Chen2018,Zhang2020} and holonomic quantum computation (NHQC)~\cite{liu2019,Sjoqvist2012,Xu2012,XuePRA2015,xue2017,Zhou2018,Hong2018PRA,Mousolou2017,Zhao2020PRA,Johansson,Zheng,Ramberg2019,jun2017} based on non-adiabatic Abelian and non-Abelian geometric phase~\cite{b1,b4} respectively have been proposed to reduce the run times of geometric quantum gates. 
Recently, non-adiabatic geometric gates have been experimentally demonstrated in different physical platforms including superconducting qubits~\cite{XXu2018prl,ZZhao2019,Song2017,Abdumalikov2013,Danilin2018,Egger2019PRAPP,Xu2018,Yan2019,Han2020}, nuclear magnetic resonance (NMR)~\cite{Feng2013,li2017,zhu2019}, and nitrogen-vacancy centers in diamond~\cite{Zu2014,Arroyo2014,s0,s2,Ishida2018,Nagata2018}, etc. However, in terms of robustness against experimental errors, the existing NGQC and NHQC gates have no sufficient preponderance over standard dynamical gates in most cases~\cite{Thomas,Zheng,Ramberg2019,jun2017}. Therefore, it is natural to ask (i) why the existing non-adiabatic geometric gates are lack of robustness against the control errors and (ii) how to maintain both the speed and the robustness of geometric gates.

In this paper, we give clear answers of the above two important issues using super robust condition proposed here. And on that basis, we demonstrated a new class of super robust non-adiabatic geometric gates which robustness against the control errors is ensured by a super robust control condition. 
We implemented our schemes in two- and three-level systems, respectively,  to realize super robust Abelian (non-Abelian) non-adiabatic geometric (holonomic) quantum gates, called SR-NGQC (SR-NHQC). 
Using the experimental parameters, numerical results indicate that our scheme can significantly improve the gate performance comparing with the existing NGQC, NHQC and standard dynamical schemes, which are in good agreement with the theoretical results. In addition, these super robust geometric quantum computation can be easily applied to various physical platforms.

\begin{figure}[tbp]
\centering\includegraphics[width=8.5cm]{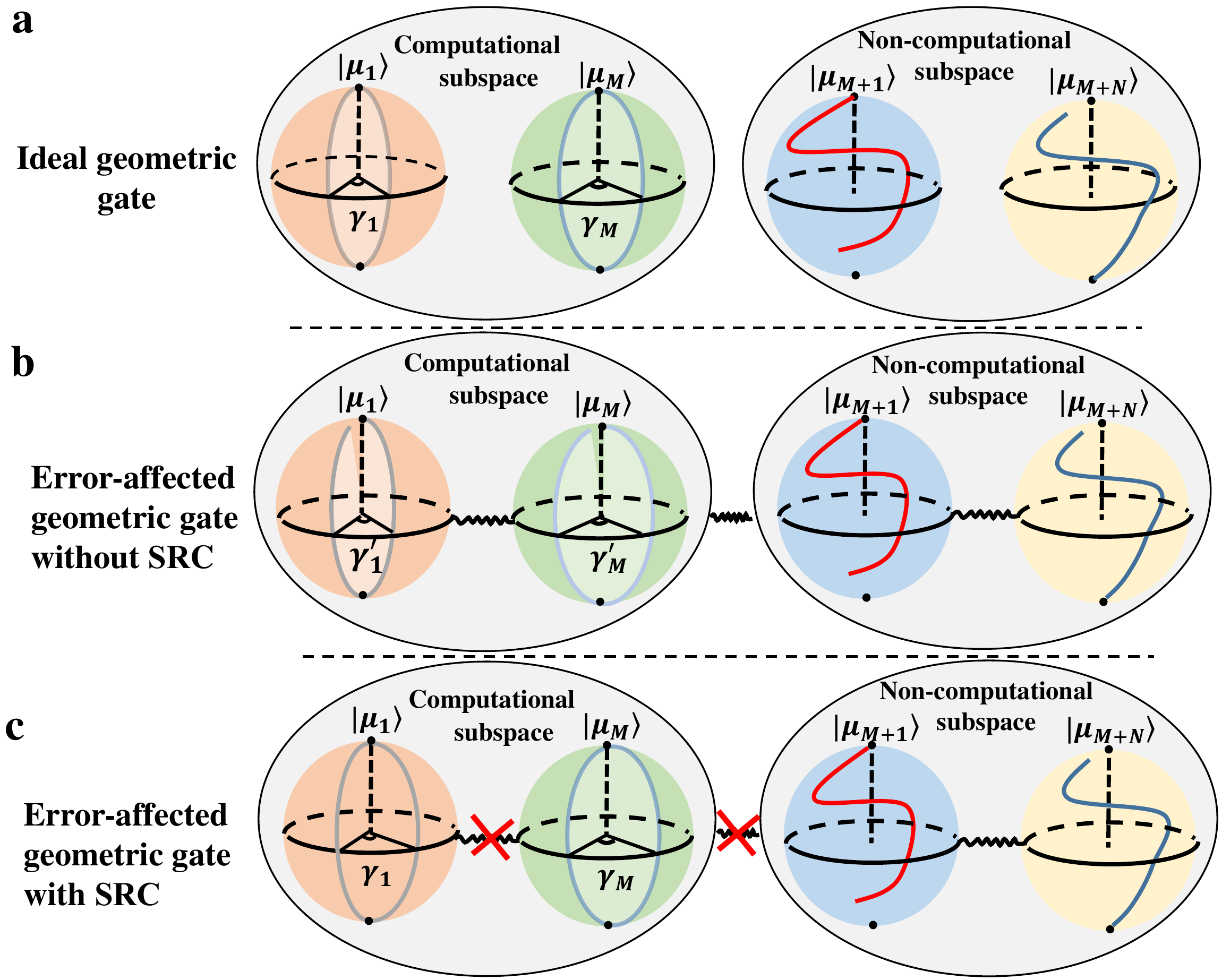}\caption{\label{setup} The illustration of our proposed implementation. (a) Conceptual explanation for the ideal non-adiabatic geometric quantum gates in a $(M+N)$-dimensional Hilbert space. The evolution state $|\mu_{k}(t)\rangle$ acquired a pure non-adiabatic geometric phase $\gamma_{k}$ under a cyclic evolution in computational subspace to construct geometric gates. Meanwhile, the state evolution in non-computational basis are completely irrelevant for geometric operations in the Hilbert space. (b) Without super-robust condition (SRC) protection, the fidelity of non-adiabatic geometric quantum gate with the presence of global control errors is limited because of the unwanted couplings of the time-dependent auxiliary states in the computational and non-computational subspace. (c) With SRC protection, the effects of the above couplings can be greatly suppressed. }
\end{figure}

\emph{General framework of  non-adiabatic geometric quantum control}.{---} Let us start with a non-degenerate quantum system described by ($\emph{M+N}$)-dimensional Hilbert space, and its evolution is governed by the Hamiltonian $H(t)$.
For any complete set of basis vectors $\left\{\left|\psi_{k}(0)\right\rangle\right\}_{k=1}^{M+N}$ at $t=0$, the time evolution operator can be written as $U\left(t\right) = {\mathcal T}{e^{ - i\int_0^t {H\left( {t'} \right)} dt'}} =
\sum\nolimits_m {\left| {{\psi _m}\left( t \right)} \right\rangle \left\langle {{\psi _m}\left( 0 \right)} \right|}$,
where the time-dependent state,  $\left| {{\psi _m}\left( t \right)} \right\rangle  = {\mathcal T}{e^{ - i\int_0^t {H\left( {t'} \right)} dt'}}\left| {{\psi _m}\left( 0 \right)} \right\rangle $, follows the Schr\"{o}dinger equation. Here, we choose a different set of time-dependent auxiliary basis states $\left\{\left|\mu_{k}(t)\right\rangle\right\}_{k=1}^{M+N}$, which makes the Hamiltonian $H(t)$ satisfy the decomposition condition.  The decomposition condition ensure that the operations in the non-computational basis $\left\{\left|\nu_{k}(t)\right\rangle\right\}_{k=M+1}^{M+N}$ are completely irrelevant for geometric operations in the computational subspace  $\left\{\left|\mu_{k}(t)\right\rangle\right\}_{k=1}^{M}$, which is given by
\begin{eqnarray}\label{Dec}
\begin{aligned}
H_{R}(t) &=V^{+}(t)\left[H(t)-i \partial_{t}\right] V(t)\\
&=H_{C}(t) \oplus H_{N}(t)  \ ,
\end{aligned}
\end{eqnarray}
where $V(t) \equiv \sum_{k}|\mu_k(t)\rangle\langle \mu_k(0)|$ and $H_{C}(t)\equiv\sum_{m, k=1}^{M}\left[\left\langle\mu_{m}\left|H(t)-i \partial_{t}\right| \mu_{k}\right\rangle\right]\Pi_{mk}(0)$ is a Hamiltonian acting on the $M$-dimensional computational subspace with $ \Pi_{mk}(0)\equiv\left|\mu_{m}(0)\right\rangle\left\langle\mu_{k}(0)\right|$, $H_{N}(t)=\sum_{m, k=M+1}^{M+N}\left[\left\langle\mu_{m}\left|H(t)-i \partial_{t}\right| \mu_{k}\right\rangle\right]\Pi_{mk}(0)$ is Hamiltonian acting on the non-computational subspace, as shown in Fig.~\ref{setup}(a). 

Now, we explain how to construct a non-adiabatic geometric gate in the computational subspace. With the help of $\left\{\left|\mu_{k}(t)\right\rangle\right\}_{k=1}^{M}$, $\left|\psi_{m}(t)\right\rangle$
can be expressed as $\left|\psi_{m}(t)\right\rangle=\sum^{M}_{l=1} C_{lm}(t)\left|\mu_{l}(t)\right\rangle$, and  the time-evolution operator in the computational subspace becomes $U_{C}(t)=\sum_{l,m=1}^{M} C_{l m}(t)\left|\mu_{l}(t)\right\rangle\left\langle\mu_{m}(0)\right|$. After a cyclic evolution, we obtain the final time evolution operator~\cite{liu2019} $U_{C}(\tau)=\sum_{m,k=1}^{M}\left[\mathcal{T} \mathrm{e}^{\mathrm{i} \int_{0}^{\tau}{\bf A}(t)+{\bf K}(t) \mathrm{d} t}\right]_{mk}\Pi_{mk}(0)$, where $\mathcal{T}$ is time ordering operator,  ${\bf A}_{lm}\equiv i\left\langle\mu_{l}(t)|\partial_{t}| \mu_{m}(t)\right\rangle$ is the matrix-valued connection one-form, and  ${\bf K}_{lm}(t) \equiv-\left\langle\mu_{l}(t)|H(t)| \mu_{m}(t)\right\rangle$ is dynamical part.

To meet the decomposition condition Eq. (\ref{Dec}), one possible set of the auxiliary state $|\mu_{k}(t)\rangle$ is found to be proportional to the time evolution states $|\psi_{k}(t)\rangle$, i.e., $|\mu_{k}(t)\rangle=e^{if_{k}(t)}|\psi_{k}(t)\rangle$. Then, we obtain the non-diagonal parts of ${\bf A}$ and ${\bf K}$ satisfy the relation
\begin{equation}\label{Nondiagaial}
{\bf A}_{lm}(t)=-{\bf K}_{lm}(t)=e^{-i\delta_{km}(t)}\left\langle\psi_{l}(t)|H(t)| \psi_{m}(t)\right\rangle \ ,
\end{equation}
where $\delta_{km}(t)=f_{k}(t)-f_{m}(t)$. \red{Using the spin-echo technique~\cite{Jones} or the pulse-shaping method~\cite{liu2019}, the accumulated dynamical phases can be erased, i.e., $\int_{0}^{\tau}K_{mm}(t) dt=0,m=1,...,M$.} In this way, we will obtain
\begin{equation}\label{GQG}
U_{C}\left( \tau\right) = \sum\limits_{m=1}^{M} {{e^{i\int_0^\tau {\bf A}_{mm}(t) dt}}\Pi_{mm}(0)} \ ,
\end{equation}
which is a non-adiabatic geometric quantum gate in the computational subspace $\left\{\left|\mu_{m}(t)\right\rangle\right\}_{m=1}^{M}$.

\emph{Super-robust condition}.{---} Now, we consider the effect of control error on the quantum evolution, the ideal Hamiltonian $H(t)$ then becomes
\begin{equation}
 H^{\prime}(t)=H(t)+\beta V(t) \ ,  
\end{equation}
where we assume that $\beta$ is a small constant, i.e., $|\beta| \ll 1$, which corresponds
to a quasi-static noise~\cite{BB1,Sunny2012}. \red{Here, $V(t)$ is the noise Hamiltonian that destroys ideal dynamics, which can be regarded as a perturbation~\cite{Ribeiro2017}.} Under this assumption, the  decomposition condition Eq. (\ref{Dec}) is broken. 
In other words,
the rotating Hamiltonian becomes $H^{\prime}_{R}(t)=\sum^{M+N}_{m=1}{\bf A}_{mm}(t)\Pi_{mm}(0)+\beta\sum^{M+N}_{m\neq k}{\bf K}_{mk}(t) \Pi_{mk}(0)$.
Then, we obtain the evolution operator as $U_{E}(\tau)=\mathcal{T}  e^{-i \int_{0}^{\tau} H^{\prime}_{R}(t) dt}$.
In general, to analytically solve this equation is difficult due to the time ordering operator.

Here, we use the Magnus expansion~\cite{Magnus1954,Blanes2009,Ribeiro2017} to perturbatively process the evolution operator $U_{E}(\tau)$. Before that, we transform to the interaction picture by defining $U_{I}(t)=\sum^{M+N}_{m=1}e^{-i\int^{t}_{0} {\bf A}_{mm}(t^{\prime})dt^{\prime}}\Pi_{mm}(0)$,
and the transformed Hamiltonian is $H^{\prime}_{IR}(t)=\beta U^{+}_{I}(t)\sum^{M+N}_{m\neq k}{\bf K}_{mk}(t) \Pi_{mk}(0)U_{I}(t)$.
The corresponding evolution operator in interaction frame is given by
\begin{equation}\label{IRU}
U_{IE}(\tau)=\mathcal{T} e^{-i \int_{0}^{\tau} H_{I R}^{\prime}(t) d t} \ .
\end{equation}
Applying Magnus expansion to the Eq. (\ref{IRU}), we have
\begin{equation}\label{Magus}
U_{IE}(\tau)=e^{\sum^{\infty}_{k=1} \Lambda_{k}(\tau)} \ ,
\end{equation}
where $\Lambda_{k}$ denotes the terms of the Magnus expansion, the two first terms of the series are given by (see, e.g.,~\cite{Magnus1954,Blanes2009,Ribeiro2017})
\begin{eqnarray}\label{MagnusExp}
\Lambda_{1}(t)&=&-i\int_{0}^{t}H_{I R}^{\prime}(t_{1}) \mathrm{d} t_{1} \ , \nonumber \\
\Lambda_{2}(t)&=&\frac{1}{2} \int_{0}^{t} \mathrm{d} t_{1} \int_{0}^{t_{1}} \mathrm{d} t_{2}\left[H_{I R}^{\prime}\left(t_{1}\right), H_{I R}^{\prime}\left(t_{2}\right)\right] \ .
\end{eqnarray}
Expanding Eq. (\ref{Magus}) in powers of $\beta$ and using Eq. (\ref{MagnusExp}), we obtain
\begin{equation}\label{EQ8}
U^{\prime}(\tau, 0)=U(\tau, 0)\left[I-i \beta D(\tau)-\frac{\beta^{2}}{2} G(\tau)+O\left(\beta^{3}\right)\right] \ ,
\end{equation}
where the element of matrix $D_{km}(t) \equiv \int_{0}^{t}\left\langle\psi_{k}(t^{\prime})|V(t^{\prime})| \psi_{m}(t^{\prime})\right\rangle dt^{\prime}$ and $G(\tau)\equiv\int_{0}^{\tau} d t\left[\dot{D}\left(t\right), D\left(t\right)\right]+D^{2}(\tau)$. \red{Consequently, the geometric gate in Eq. \eqref{GQG} under the control error becomes
$U_{C}^{\prime}(\tau,0)=\sum^{M}\nolimits_{m=1} {\left| {{\psi^{\prime}_m}\left( \tau \right)} \right\rangle \left\langle {{\psi _m}\left( 0 \right)} \right|}=
\sum_{m=1}^{M}N_{rm}U^{\prime}(\tau,0)\Pi_{mm}(0)$, where $N_{rm}=1/\sqrt{1+\beta^{2}\sum^{M+N}_{k=1}|D_{km}|^{2}}$ is the state normalized coefficient.} To further evaluate the performance of the quantum operation caused by the control error, the gate fidelity~\cite{Souza2011,Genov2017} is taken by
\begin{eqnarray}\label{Fidelity}
\begin{aligned}
F&=\frac{1}{M}\left|\operatorname{Tr}\left(U^{\prime}_{C} U_{C}^{\dagger}\right)\right| \\
&\approx1-\frac{\beta{^2}}{2M}\sum^{M}_{m=1}\sum_{k=1}^{M+N}\left|D_{km}\right|^{2}-\mathcal{O}(\beta^{4}) \ .
\end{aligned}
\end{eqnarray}
Consequently, we can achieve  a super robust gates with the fourth-order  error  dependence  against  the  control  error, as long as the following  condition is satisfied,
\begin{equation}\label{GPC2}
D_{km}\equiv\int^{\tau}_{0}\left\langle\psi_{k}(t)|V(t)|\psi_{m}(t)\right\rangle dt=0 \ ,
\end{equation}
where $k=1,...,M+N$ and $m=1,2...,M$. 

\begin{figure}[tbp]
\centering
\includegraphics[width=8.5cm]{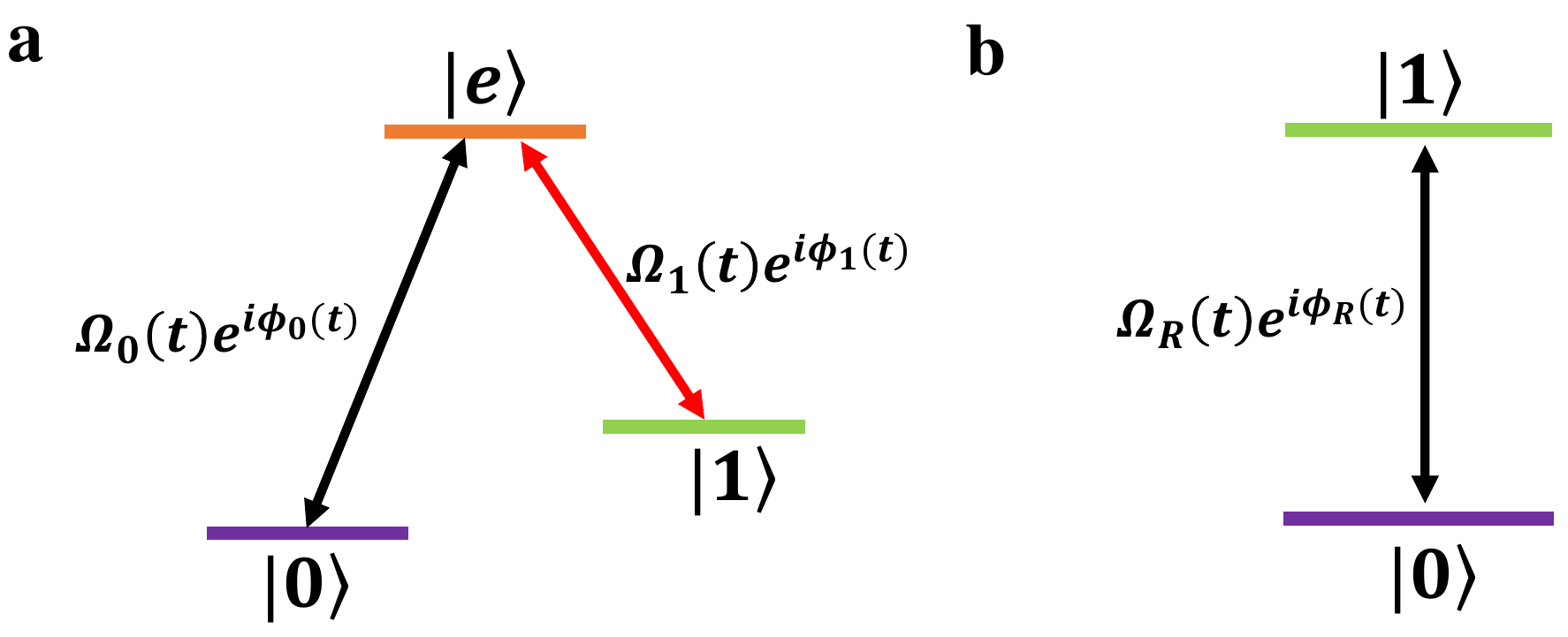}
\caption{The level structure and coupling configuration for the construction of SR-NHQC and SR-NGQC gates. (a) The driven pulses with amplitudes $\Omega_{0}$ and $\Omega_{1}$ resonantly couple $|0\rangle$ and $|1\rangle$ to $|e\rangle$ with the phases $\phi_{0}$ and $\phi_{1}$, respectively.  (b) The driven pulse with amplitudes $\Omega_{R}$ resonantly couples $|0\rangle$ and to $|1\rangle$ with the time-independent phase $\phi_{R}$.}\label{single}
\end{figure}

As a demonstration of Eq. (\ref{GPC2}), we consider that the control error is global with $V(t)=H(t)$. In this case, the super robust condition can be expressed by 
\begin{equation}\label{GPC}
D^{\prime}_{km}=\int^{\tau}_{0}\left\langle\psi_{k}(t)|H(t)|\psi_{m}(t)\right\rangle dt=0 \ .
\end{equation}
In this way, the effects of global control errors can be greatly suppressed. Here, we further illustrate the geometric meaning of Eq. (\ref{GPC}): $D^{\prime}_{mm}=0$ with $k=m$ erases the accumulated dynamical phases; $D_{km}=\int^{\tau}_{0}e^{i\delta_{km}}{\bf A}_{km} dt=0$ with $k\neq m$ suppresses both couplings between the time-dependent auxiliary states $|\mu_{m}(t)\rangle$ and $|\mu_{n}(t)\rangle$ in the computational subspace and non-computational subspace under the control errors, as shown in Fig. \ref{setup}(b) and \ref{setup}(c).

Note that the key difference between the previous NGQC, NHQC schemes and the super robust condition-based scheme proposed here is that the Hamiltonians have different constraints.
In the NGQC case, the Hamiltonian is only required to satisfy the constraint $D^{\prime}_{mn}=0$.  The constraint of NHQC is set as $D^{\prime}_{mn}=0$ with $m,n=1,...,M$. We can clearly see that the constrains of NGQC and NHQC are necessary not sufficient condition of super robust gates. Therefore we can explain  why some previous NGQC and NHQC lack robustness to the global control errors.

Furthermore, we can further explain the origin of the robustness of adiabatic GQC~\cite{Jones,Wu2013PRA,Huang2019} against the control errors. The adiabatic condition~\cite{b1,Kolodrubetz2017} for non-degenerate system is $\left|A_{km}\right|=\left|\frac{\left\langle e_{k}\left|\partial_{t} H\right| e_{m}\right\rangle}{E_{k}-E_{m}}\right| \ll 1$ for $k\neq m$, where $\{E_{k}(t)\}$ and $|e_{k}(t)\rangle$ denote the instantaneous eigenvalues
and eigenvectors of $H(t)$. In this case, we consider the auxiliary basis states $|\mu_{k}(t)\rangle$ to be identical to the eigenvectors $|e_{k}(t)\rangle$. The adiabatic condition ensures that the conditions of Eq. (\ref{Dec}) and Eq. (\ref{GPC}) are met. Consequently, we can be verified that the fidelity of adiabatic GQC is at least fourth-order error dependence against global control errors using the Eq. (\ref{Fidelity}).  In the following, to illustrate the working mechanism of super robust geometric gates, we shall give two simple examples of SR-NGQC and SR-NHQC for two- and three-level quantum systems,respectively

\emph{Example 1: SR-NHQC}.{---} The main idea of previous NHQC is to generate a non-adiabatic non-Abelian geometric gate in a three-level system, as shown in  Fig. \ref{single} (a).
Under the rotating-wave approximation, the system Hamiltonian is given by: $H(t) = \sum_{i=0}^{1}\frac{1}{2}\left[\Omega_i(t) e^{i\phi_i }| i\rangle\langle e|+ {h.c.} \right]$.
We define a bright state, $| b\rangle \equiv \sin(\frac{\theta}{2})e^{i\phi }| 0\rangle + \cos(\frac{\theta}{2})|1\rangle$, where $\phi \equiv \phi_0(t)-\phi_1(t)$ and $\tan(\theta/2) \equiv \Omega_0(t)/\Omega_1(t)$.
We shall keep $\theta$ and $\phi$, hence $\left| b \right\rangle$ to be time independent. The Hamiltonian $H(t)$ can then be rewritten as: $H(t) =\frac{1}{2}(\Omega(t)e^{-i\phi_1(t)}|b\rangle \langle e| +h.c.)$,
where $\Omega(t) \equiv \sqrt{\Omega_0(t)^2+\Omega_1(t)^2}$ is the Rabi frequency of $H(t)$.

\begin{figure}[htbp]
\centering
\includegraphics[width=8.5cm]{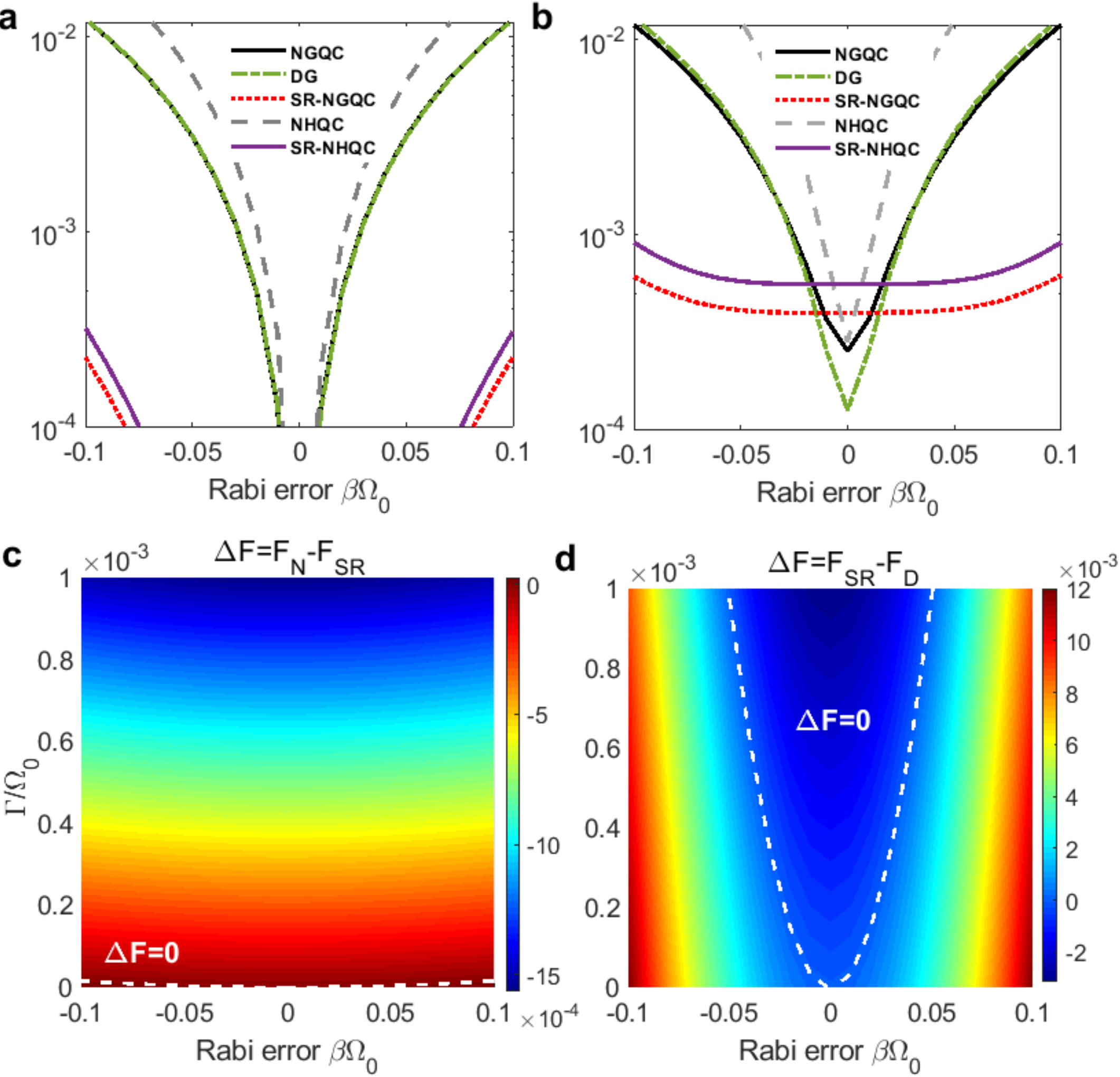}
\caption{Numerical robustness comparison of the NOT gate with various approaches.  The NOT gate infidelities $1-F$ of DG, NGQC, SR-NGQC, NHQC, and SR-NHQC are set as a function of control error of Rabi frequency, i.e., the relative pulse deviation $\beta$ (a) without and (b) with the decoherence. The NOT gate fidelities difference $\Delta F=F_{G}-F_{D}$ ($\Delta F=F_{N}-F_{D}$) (d) [(c)] between SR-NGQC (NGQC) and DG are set as the decoherence parameter $\Gamma$ and the relative pulse deviation $\beta$.  }\label{Robust}
\end{figure}

\red{Here, we choose the time-dependent auxiliary basis states as $\left|\mu_{1}(t)\right\rangle=|d\rangle$, $\left|\mu_{2}(t)\right\rangle=\cos \frac{\alpha(t)}{2}|b\rangle-i \sin \frac{\alpha(t)}{2} e^{-i \phi_{1}(t)}|e\rangle$ and $\left|\mu_{3}(t)\right\rangle=-i\sin \frac{\alpha(t)}{2}e^{i\phi_{1}(t)}|b\rangle+
\cos\frac{\alpha(t)}{2}|e\rangle$ with $\alpha(t)=\int_{0}^{t}\Omega(t^{\prime}) dt^{\prime}$ (see
Supplemental Material~\cite{SM} for details). Therefore, the rotating Hamiltonian Eq. (\ref{Dec}) in the auxiliary basis states is given by
$H_{R}(t)=\frac{1-\cos \alpha}{2}\dot{\phi}_{1}|b\rangle\langle b|\oplus \frac{\cos \alpha -1}{2}\dot{\phi}_{1}| e\rangle\langle e|$. The corresponding computational Hamiltonian  and
non-computational Hamiltonian are given by $H_{C}(t)=\frac{1-\cos \alpha}{2} \dot{\phi}_{1}|b\rangle\langle b|+0|d\rangle\langle d|$
and $H_{N}(t)=\frac{\cos \alpha -1}{2} \dot{\phi}_{1}| e\rangle\langle e|$, respectively.} Since the  parallel transport conditions of previous NHQC~\cite{b5,b6,Thomas,zhao2017,Li2020PRR,Chen2018,Zhang2020,XXu2018prl,ZZhao2019,Song2017}, i.e., $\left\langle\psi_{m}(t)|H(t)|\psi_{n}(t)\right\rangle=0$ ( $n,m=1,2$) is satisfied, the super robust condition Eq. (\ref{GPC}) can be reduced to,
\begin{equation}\label{GPCThree}
  D^{\prime}_{23}=\int_{0}^{\tau} \frac{\dot{\alpha}}{2}\exp(-i\int^{t}_{0}\dot{\phi}_{1}/\cos\alpha dt') dt=0 \ .
\end{equation}
Consequently, we obtain the following unitary transformation matrix in the basis states $\{|\mu_{1}(0)\rangle,|\mu_{2}(0)\rangle\}$, i.e., $U_{C}(\tau)=e^{i\gamma_{g}}|b\rangle\langle b|+|d\rangle\langle d|$,
where $\gamma_{g}=\int^{\tau}_{0}\frac{\cos \alpha -1}{2} \dot{\phi}_{1}dt=\frac{1}{2}\int^{\phi_{1}(\tau)}_{\phi_{1}(0)}\int^{\alpha(\tau)}_{\alpha(0)}\sin\alpha  d\alpha d\phi_{1}$, which shows that the geometric
phase $\gamma_{g}$ exactly equals to half of the solid angle.
Note that the non-adiabatic holonomic gate can be spanned by the logical basis $\{|0\rangle,|1\rangle\}$, i.e.,
\begin{equation}\label{GPCBased}
U (\gamma_{g}, \theta, \phi)=e^{i\frac{\gamma_{g}}{2} {\bf n} \cdot {\bf \sigma}}
\end{equation}
where $\bm{n}=(\sin\theta\cos\phi,\sin\theta\sin\phi,\cos\theta)$, $\bm{\sigma}$ are the Pauli matrices.  Eq. (\ref{GPCBased}) describes a rotational operation around the $\bm{n}$ axis by a $\gamma_{g}$ angle,
ignoring a global phase factor $e^{-i\frac{\gamma_{g}}{2}}$.  Since both $\bm{n}$ and $\gamma_{g}$ can take any value,  Eq. (\ref{GPCBased}) denotes a set of universal single-qubit gates in the qubit subspace.

\emph{Example 2:  SR-NGQC}.{---} As an another example, one can also apply our scheme to a two-level system, as shown in  Fig. \ref{single}(b).
The system Hamiltonian can be written as: $H_1(t) = \frac{\Omega_{R}(t)}{2} e^{i\phi_R(t) }| 0\rangle\langle 1|+ {h.c.}$. The time-dependent auxiliary basis states are taken by   $\left|\zeta_{1}(t)\right\rangle=\cos \frac{\alpha_{R}(t)}{2}|0\rangle-
 i \sin \frac{\alpha_{R}(t)}{2} e^{-i \phi_{R}(t)}|1\rangle$ and $\left|\zeta_{2}(t)\right\rangle=-i\sin \frac{\alpha_{R}(t)}{2}e^{i \phi_{R}(t)}|0\rangle+
\cos\frac{\alpha_{R}(t)}{2}|1\rangle$ with $\alpha_{R}(t)=\int^{t}_{0}\Omega_{R}(t^{\prime}) dt^{\prime}+\alpha_{R}(0)$. Under these settings, the condition of previous NGQC~\cite{b5,b6,Thomas,zhao2017,Li2020PRR,Chen2018,Zhang2020,XXu2018prl,ZZhao2019,Song2017}, i.e.,
$D^{\prime}_{mm}=\int^{\tau}_{0}{\bf K}_{mm}(t) dt=0, m=1,2$, is satisfied. Thus, the super robust condition Eq. (\ref{GPC}) for NGQC becomes
\begin{equation}\label{GPCTwo}
  D^{\prime}_{12}=\int_{0}^{\tau} \frac{\dot{\alpha}_{R}}{2}\exp(-i\int^{t}_{0}\dot{\phi}_{R}/\cos\alpha_{R} dt') dt=0 \ .
\end{equation}
Similar to the SR-based NHQC case, we can also obtain the universal single-qubit gates in Eq. (\ref{GPCBased}), i.e.,
$U \left[\gamma_{R}, \alpha_{R}(0), \phi_{R}(0)\right]$ with the geometric phase $\gamma_{R}=\int^{\tau}_{0}\dot{\phi}_{R}\left(\cos \alpha_{R} -1\right)/2 dt$. 

\emph{Numerical simulations}.{---} To investigate the noise-resilient feature of super-robust geometric gates against the global control error, we take the NOT gate as a typical example to compare the performances of SR-NHQC and SR-NGQC approaches with that of the related standard dynamical gate (DG), NGQC and NHQC approaches. Before that, we choose time-independent Rabi frequencies of all the gates and keep the same as $\Omega(t)=\Omega_{R}(t)=\Omega_{0}$, and the corresponding gate times are sketched in Table \ref{table1} (see Supplemental  Material~\cite{SM} for  details for).  Here, we assume the relative pulse deviation $\beta$ of driven pulse to vary in the range of $\beta\in [-0.1,0.1]$. As shown in Fig. \ref{Robust}(a), SR-NHQC and SR-NGQC are always more robust than DG, NHQC and NGQC gate without the consideration of decoherence. The numerical results are in good agreement with the theoretical results, as shown in Table \ref{table1}.

In fact, the decoherence process is unavoidable. To evaluate the performance of the gates with the consideration the influence of decoherence, the Lindblad master  equation~\cite{Lindblad} is used here. To be more eloquent, we set the parameters from the the current experiments~\cite{Buluta2011,chen2016,Krantz2016}: the decay and dephasing rates of the qubit are taken as $\Gamma_{1}=\Gamma_{2}
=\Gamma=10^{-4} \Omega_{0}$ in our numerical simulation. From the Fig. \ref{Robust}(b), with the consideration of both the Rabi control error and the decoherence effect, we find the SR-NGQC can significantly improve the the robustness of quantum gates with the relative pulse deviation $|\beta|>0.01$.

\begin{table}[tp]
\centering \caption{The robustness comparison with the NOT gate of standard dynamical gate (DG), non-adiabatic geometric gate, holonomic gate and super robust geometric quantum gates.}
\begin{tabular}{cccccc}
\hline
 Types & System level & Gate time & Fidelity  & References \\
   \hline
DG  & 2&$\pi/\Omega_{0}$ &  $1-\beta^{2}\pi^2/8$ & \\
NGQC  & 2 & $2\pi/\Omega_{0}$   & $1-\beta^{2}\pi^2/8$ & \cite{b5,b6,Thomas,zhao2017,Li2020PRR,Chen2018,Zhang2020,XXu2018prl,ZZhao2019,Song2017} \\
SR-NGQC  & 2 & $3\pi/\Omega_{0}$  & $1-\mathcal{O}(\beta^{4})$ & This work\\
NHQC  & 3 & $2\pi/\Omega_{0}$  & $1-\beta^{2}\pi^2/3$ & \cite{Sjoqvist2012,Xu2012,XuePRA2015,xue2017,Zhou2018,Hong2018PRA,Mousolou2017,Zhao2020PRA,Johansson,Zheng,Ramberg2019,jun2017,Abdumalikov2013,Xu2018,Feng2013,li2017,
zhu2019,Zu2014,Arroyo2014,s0,s2,Ishida2018,Nagata2018,Danilin2018,Egger2019PRAPP} \\
SR-NHQC  & 3 & $4\pi/\Omega_{0}$   & $1-\mathcal{O}(\beta^{4})$ & This work \\
 \hline
\end{tabular}\label{table1}
\end{table}

On other hand, one can find from Fig.~\ref{Robust}(b) that
the DG is better than NGQC and SR-NGQC for the relative pulse deviation $\beta\in [-0.01,0.01]$, since decoherence is the main factor in this case.
In order to balance between the decoherence and the Rabi control errors, we plot the NOT gate fidelities difference $\Delta F=F_{G}-F_{D}$ ($\Delta F=F_{N}-F_{D}$) between SR-NGQC (NGQC) and DG as the decoherence parameter $\Gamma$ and the relative pulse deviation $\beta$, as shown in Fig. \ref{Robust}(d) [\ref{Robust}(c)]. We find that the best scheme in this case depends on the relative importance between the decoherence and the Rabi control error. An open question for further development,  whether designing hybrid SR-NGQC and DG scheme is optimal for quantum control. In addition, we clearly know the NOT gate of NGQC has no particular advantage compared to DG with both above errors, as shown in the Fig. \ref{Robust}(c).

\emph{Conclusion and outlook}.{---}In conclusion, we have explained why the existing non-adiabatic geometric gates are so sensitive to the control errors, but also proposed the scheme of super-robust non-adiabatic geometric gates. Specifically, we take two examples in two- and three-level systems for realizing super robust condition-based Abelian (non-Abelian) non-adiabatic geometric (holonomic) quantum gates, respectively. The theoretical and numerical results indicate that our scheme can significantly improve the gate performance comparing with the existing geometric and standard dynamical schemes with the experimental parameters. Moreover, our scheme can be also extended to construct two-qubit geometric gates~\cite{XXu2018prl, Egger2019PRAPP,Zu2014,Nagata2018} to realize a universal SR-based geometric gate set. In addition, this extensible approach of SR-based geometric gates can be applied to various physical platforms such as superconducting circuits, quantum dots and trapped ions. For future work, it would be attractive for fault-tolerant quantum computation to combine SR-based geometric scheme with the decoherence-free subspace (DFS)~\cite{Xu2012,xue2017,XuePRA2015,Song2016,liang2014,Feng2009} encoding model (Surface codes~\cite{Bravyi,Dennis2002,Fowler2012,ZhangJ2018,Wu2020}) to further suppress the dephasing noises (local errors).  In addition, it would also be interesting to further optimize the magnetic field
sensitivity and maximum field range of geometric-phase magnetometry~\cite{NC2018} via our SR-NGQC scheme.

\bigskip

\acknowledgments

The authors thank Prof. S.-L. Su for helpful discussions and improving the paper. This work is supported by the Natural Science Foundation of Guangdong Province (Grant No. 2017B030308003), the Key R \& D Program of Guangdong province (Grant No. 2018B030326001), the Science, Technology and Innovation Commission of Shenzhen Municipality (Grant No. JCYJ20170412152620376 and No. JCYJ20170817105046702 and No. KYTDPT20181011104202253), National Natural Science Foundation of China (Grant No. 11875160 and No. U1801661), the Economy, Trade and Information Commission of Shenzhen Municipality (Grant No. 201901161512), Guangdong Provincial Key Laboratory (Grant No.2019B121203002).


\begin{thebibliography}{99}

\bibitem{b2}
M. V. Berry, Quantal phase factors accompanying adiabatic changes, Proc. R. Soc. Lond. A $\textbf{392}$, 45 (1984).

\bibitem{b1}
Y. Aharonov, and J. Anandan, Phase change during a cyclic quantum evolution, Phys. Rev. Lett. $\textbf{58}$, 1593 (1987).

\bibitem{Zanardi1999}
P. Zanardi, and M. Rasetti, Holonomic quantum computation, Phys. Lett. A \textbf{264}, 94 (1999).

\bibitem{b4}
J. Anandan, Non-adiabatic non-abelian geometric phase, Phys. Lett. A \textbf{133}, 171 (1988).

\bibitem{gqc}
E. Sj\"{o}qvist, Trend: A new phase in quantum computation, Physics $\textbf{1}$, 35 (2008).


\bibitem{Zhu2005}
S.-L. Zhu, and P. Zanardi, Geometric quantum gates that are robust against stochastic control errors, Phys. Rev. A $\textbf{72}$, 020301(R) (2005).

\bibitem{Berger2013} S. Berger, M. Pechal, A. A. Abdumalikov, Jr. C. Eichler, L. Steffen, A. Fedorov, A. Wallraff, and S. Filipp, Exploring the effect of noise on the Berry phase, Phys. Rev. A $\textbf{87}$,
060303(R) (2013).

\bibitem{Chiara}
G. D. Chiara and G. M. Palma, Berry phase for a spin-1/2 particle in a classical fluctuating field, Phys. Rev. Lett. $\textbf{91}$, 090404 (2003).

\bibitem{Leek}
P. J. Leek, J. M. Fink, A. Blais, R. Bianchetti, M. G\"{o}ppl, J. M. Gambetta, D. I. Schuster, L. Frunzio, R. J. Schoelkopf, and A. Wallraff, Observation of Berry's phase in a solid state qubit, Science $\textbf{318}$, 1889 (2007).  

\bibitem{Filipp}
S. Filipp, J. Klepp, Y. Hasegawa, C. Plonka-Spehr, U. Schmidt, P. Geltenbort, and H. Rauch, Experimental demonstration of the stability of Berry's phase for a spin-1/2 particle, Phys. Rev. Lett. $\textbf{102}$, 030404 (2009).
%
\bibitem{b3}
F. Wilczek, and A. Zee, Appearance of gauge structure in simple dynamical systems, Phys. Rev. Lett. $\textbf{52}$, 2111 (1984).

\bibitem{Jones}
J. A. Jones, V. Vedral, A. Ekert, and G. Castagnoli, Geometric quantum computation using nuclear magnetic resonance, Nature $\textbf{403}$, 869 (2000). 

\bibitem{Wu2013PRA} H. Wu, E. M. Gauger, R.E. George, M. Mottonen, H. Riemann, N. V. Abrosimov, P. Becker, H. Pohl, K. M. Itoh, M. L.W. Thewalt, and J. J. L. Morton, Geometric phase gates with adiabatic control in electron spin resonance, Phys. Rev. A {\bf 87}, 032326 (2013).

\bibitem{Huang2019} Y.-Y. Huang, Y.-K. Wu, F. Wang, P.-Y. Hou,W.-B. Wang, W.-G. Zhang, W.-Q. Lian, Y.-Q. Liu, H.-Y. Wang, H.-Y. Zhang, \emph{et al}., 
Experimental Realization of Robust Geometric Quantum Gates with Solid-State Spins, Phys. Rev. Lett. {\bf 122}, 010503 (2019).

\bibitem{Duan2001a}
L. M. Duan, J. I. Cirac and P. Zoller, Geometric manipulation of trapped ions for quantum computation, Science $\textbf{292}$, 1695 (2001).  

\bibitem{lian2005}
L.-A. Wu, P. Zanardi, and D. A. Lidar, Holonomic Quantum Computation in Decoherence-Free Subspaces, Phys. Rev. Lett. $\textbf{95}$, 130501 (2005).

\bibitem{b5} Wang, X. B., and M. Keiji, Nonadiabatic conditional geometric phase shift with NMR, Phys. Rev. Lett. $\textbf{87}$, 097901 (2001).

\bibitem{b6} S.-L. Zhu, and Z. D. Wang, Implementation of universal quantum gates based on nonadiabatic geometric phases, Phys. Rev. Lett. $\textbf{89}$, 097902 (2002).

\bibitem{Thomas} J. T. Thomas, M. Lababidi, and M. Tian, Robustness of single-qubit geometric gate against systematic error, Phys. Rev. A $\textbf{84}$, 042335 (2011).

\bibitem{zhao2017} P. Z. Zhao, X.-D. Cui, G. F. Xu, E. Sj\"oqvist, and D. M. Tong, Rydberg-atom-based scheme of non-adiabatic geometric quantum computation, Phys. Rev. A {\bf 96}, 052316 (2017).

\bibitem{Li2020PRR} K. Z. Li, P. Z. Zhao, and D. M. Tong, Approach to realizing non-adiabatic geometric gates with prescribed evolution paths, Phys. Rev. Research {\bf 2}, 023295 (2020).

\bibitem{Chen2018} T. Chen and Z.-Y. Xue, non-adiabatic geometric quantum computation with parametrically tunable coupling, Phys. Rev. Appl. {\bf 10}, 054051 (2018).

\bibitem{Zhang2020} C. Zhang, T. Chen, S. Li and Z.-Y. Xue, High-fidelity geometric gate for silicon-based spin qubits, Phys. Rev. A {\bf 101}, 052302 (2020).


\bibitem{liu2019} B.-J. Liu, X.-K. Song, Z.-Y. Xue, X. Wang, and M.-H. Yung, Plug-and-Play Approach to nonadiabatic Geometric Quantum Gates, Phys. Rev. Lett. \textbf{123}, 100501 (2019).


\bibitem{Sjoqvist2012}
E. Sj\"{o}qvist, D. M. Tong, L. M. Andersson, B. Hessmo, M. Johansson, and K. Singh, Non-adiabatic holonomic quantum computation, New J. Phys. $\textbf{14}$, 103035 (2012).

\bibitem{Xu2012} G. F. Xu, J. Zhang, D. M. Tong, E. Sj\"{o}qvist, and L. C. Kwek, Nonadiabatic Holonomic Quantum Computation in Decoherence-free Subspaces, Phys. Rev. Lett. $\textbf{109}$,170501 (2012).

\bibitem{XuePRA2015} Z.-Y. Xue, J. Zhou, and Z. D. Wang, Universal holonomic quantum gates in decoherence-free subspace on superconducting circuits,  Phys. Rev. A {\bf 92}, 022320 (2015).

\bibitem{xue2017}
Z.-Y. Xue, F.-L. Gu, Z.-P. Hong, Z.-H. Yang, D.-W. Zhang, Y. Hu, and J. Q. You, non-adiabatic holonomic quantum computation with dressed-state qubits, Phys. Rev. Appl. $\textbf{7}$, 054022 (2017).

\bibitem{Zhou2018} J. Zhou, B. J. Liu, Z. P. Hong, and Z. Y. Xue, Fast holonomic quantum computation based on solid-state spins with all-optical control, Sci. China-Phys. Mech. Astron. {\bf 61}, 010312 (2018).

\bibitem{Hong2018PRA} Z.-P. Hong, B.-J. Liu, J.-Q. Cai, X.-D. Zhang, Y. Hu, Z. D. Wang, and Z.-Y. Xue, Implementing universal non-adiabatic holonomic quantum gates with transmons, Phys. Rev. A {\bf 97}, 022332 (2018).

\bibitem{Mousolou2017} V. Azimi Mousolou, “Electric non-adiabatic geometric entangling gates on spin qubits," Phys. Rev. A {\bf 96}, 012307 (2017).

\bibitem{Zhao2020PRA} P. Z. Zhao, K. Z. Li, G. F. Xu, and D. M. Tong, “General approach for constructing Hamiltonians for non-adiabatic holonomic quantum computation," Phys. Rev. A {\bf 101}, 062306 (2020).

\bibitem{Johansson}  M. Johansson, E. Sj\"{o}qvist, L. M. Andersson, M. Ericsson, B. Hessmo, K. Singh, and D. M. Tong, Robustness of non-adiabatic holonomic gates, Phys. Rev. A
$\textbf{86}$, 062322 (2012).

\bibitem{Zheng}
S. B. Zheng, C. P. Yang, and F. Nori, Comparison of the sensitivity to systematic errors between nonadiabatic non-Abelian geometric gates and their dynamical counterparts, Phys. Rev. A $\textbf{87}$, 032326 (2016).

\bibitem{Ramberg2019} N. Ramberg and E. Sj\"{o}qvist, Environment-Assisted Holonomic Quantum Maps, Phys. Rev. Lett. {\bf 122}, 140501 (2019).

\bibitem{jun2017}
J. Jing, C.-H. Lam, and L.-A. Wu, Non-Abelian holonomic transformation in the presence of classical noise, Phys. Rev. A $\textbf{95}$, 012334 (2017).

\bibitem{XXu2018prl} Y. Xu, Z. Hua, Tao Chen, X. Pan, X. Li, J. Han, W. Cai, Y. Ma, H. Wang, Y. P. Song, Z.-Y. Xue, and L. Sun, Experimental implementation of universal non-adiabatic geometric quantum gates in a superconducting circuit, Phys. Rev. Lett. {\bf 124}, 230503 (2020).

\bibitem{ZZhao2019} P. Z. Zhao, Z. Dong, Z. Zhang, G. Guo, D. M. Tong, and Y. Yin, Experimental realization of non-adiabatic geometric gates with a superconducting xmon qubit, arXiv: 1909.09970 (2019).

\bibitem{Song2017} C. Song, S.-B. Zheng, P. Zhang, K. Xu, L. Zhang, Q. Guo, W. Liu, D. Xu, H. Deng, K. Huang, \emph{et al}., Continuous-variable geometric phase and its manipulation for quantum computation in a superconducting circuit, Nat. Commun. {\bf 8}, 1061 (2017).

\bibitem{Yan2019} T. Yan, B.-J. Liu, K. Xu, C. Song, S. Liu, Z. Zhang, H. Deng, Z. Yan, H. Rong, M.-H. Yung, Y. Chen, and D. Yu, Experimental realization of non-adiabatic shortcut to non-Abelian geometric gates, Phys. Rev. Lett. {\bf 122}, 080501 (2019).

\bibitem{Han2020} Z. Han, Y. Dong, B. Liu, X. Yang, S. Song, L. Qiu, D. Li, J. Chu, W. Zheng, J. Xu, T. Huang, Z. Wang, X. Yu, X. Tan, D. Lan, M.-H. Yung, Y. Yu, Experimental Realization of Universal Time-optimal non-Abelian Geometric Gates, arXiv:2004.10364 (2020).

\bibitem{Abdumalikov2013}
A. A. Abdumalikov, J. M. Fink, K. Juliusson, M. Pechal, S. Berger, A. Wallraff, and S. Filipp, Experimental realization of non-Abelian non-adiabatic geometric gates, Nature $\textbf{496}$, 482-485 (2013).

\bibitem{Xu2018} Y. Xu, W. Cai, Y. Ma, X. Mu, L. Hu, T. Chen, H. Wang, Y. P. Song, Z.-Y. Xue, Z.-Q. Yin, and L. Sun, Single-Loop Realization of Arbitrary Nonadiabatic Holonomic Single Qubit Quantum Gates in a Superconducting Circuit, Phys. Rev. Lett. {\bf 121}, 110501 (2018).

\bibitem{Danilin2018} S. Danilin, A. Veps\"{a}l\"{a}inen, G. S. Paraoanu, Experimental state control by fast non-Abelian holonomic gates with a superconducting qutrit,
Physica Scripta {\bf 93} (5), 055101 (2018).


\bibitem{Egger2019PRAPP} D. J. Egger, M. Ganzhorn, G. Salis, A. Fuhrer, P. Muller, P. K. Barkoutsos, N. Moll, I. Tavernelli, and S. Filipp, Entanglement Generation in Superconducting Qubits Using Holonomic Operations,
 Phys. Rev. Appl. {\bf 11}, 014017 (2019).
 
\bibitem{Feng2013}
G. Feng, G. Xu, and G. Long, Experimental Realization of non-adiabatic Holonomic Quantum Computation, Phys. Rev. Lett. {\bf 110}, 190501(2013).

\bibitem{li2017} H. Li, L. Yang, and G. Long, Experimental realization of single-shot non-adiabatic holonomic gates in nuclear spins, Sci. China: Phys., Mech. Astron. {\bf 60}, 080311(2017).

\bibitem{zhu2019}
Z. Zhu, T. Chen, X. Yang, J. Bian, Z.-Y. Xue, and X. Peng, Single-Loop and Composite-Loop Realization of non-adiabatic Holonomic Quantum Gates in a Decoherence-Free Subspace, Phys. Rev. Appl. {\bf 12}, 024024 (2019).

\bibitem{Zu2014} C. Zu, W.-B. Wang, L. He, W.-G. Zhang, C.-Y. Dai, F. Wang and L.-M. Duan, Experimental realization of universal geometric quantum gates with solid-state spins, Nature (London) {\bf 514}, 72 (2014).

\bibitem{Nagata2018} K. Nagata, K. Kuramitani, Y. Sekiguchi, and H. Kosaka, Universal holonomic quantum gates over geometric spin qubits with polarised microwaves, Nat. Commun. {\bf 9}, 3227 (2018).

\bibitem{Arroyo2014} S. Arroyo-Camejo, A. Lazariev, S. W. Hell, and G. Balasubramanian, Room temperature high-fidelity holonomic single-qubit gate on a solid-state spin, Nat. Commun. {\bf 5}, 4870 (2014).

\bibitem{s0}Y. Sekiguchi, N. Niikura, R. Kuroiwa, H. Kano, and H. Kosaka, Optical holonomic single quantum gates with a geometric spin under a zero field,  Nat. Photonics  $\textbf{11}$, 309 (2017). 
%
\bibitem{s2} B. B. Zhou, P. C. Jerger, V. O. Shkolnikov, F. Joseph Heremans, G. Burkard, and D. D. Awschalom, Holonomic Quantum Control by Coherent Optical Excitation in Diamond,
Phys. Rev. Lett. $\textbf{119}$, 140503 (2017).

\bibitem{Ishida2018} N. Ishida, T. Nakamura, T. Tanaka, S. Mishima, H. Kano, R. Kuroiwa, Y. Sekiguchi, and H. Kosaka, Universal holonomic single quantum gates over a geometric spin with phase-modulated polarized light, Opt. Lett. {\bf 43}, 2380 (2018).

\bibitem{Kolodrubetz2017} M. Kolodrubetz, D. Sels, P. Mehta, and A. Polkovnikov, Geometry and non-adiabatic response in quantum and classical systems, Phys. Rep. {\bf 697}, 1 (2017).


\bibitem{BB1}
X. Rong, J. Geng, F. Shi, Y. Liu, K. Xu, W. Ma, F. Kong, Z. Jiang, Y. Wu, and J. Du, Experimental fault-tolerant universal quantum gates with solid-state spins under ambient conditions, Nat. Commun. $\textbf{6}$ (2015).

\bibitem{Sunny2012} X. Wang, L. S. Bishop, J. P. Kestner, E. Barnes, K. Sun, and S. D. Sarma, Composite pulses for robust universal control of singlet–triplet qubits, Nat. Commun. {\bf 3}, 997 (2012).

\bibitem{Ribeiro2017} H. Ribeiro, A. Baksic, and A. A. Clerk, Systematic magnus-based approach for suppressing leakage and non-adiabatic errors in quantum dynamics, Phys. Rev. X {\bf 7}, 011021 (2017).


\bibitem{Magnus1954} W. Magnus, On the Exponential Solution of Differential Equations for a Linear Operator,  Commun. Pure Appl. Math. {\bf 7}, 649 (1954).

\bibitem{Blanes2009} S. Blanes, F. Casas, J. A. Oteo, and J. Ros, The Magnus Expansion and Some of Its Applications,  Phys. Rep. {\bf 470}, 151 (2009).
\bibitem{Souza2011} A. Souza, G. A. Alvarez, and D. Suter, Robust dynamical decoupling, Phil. Trans. R. Soc. A {\bf 370}, 4748-4769 (2012).

\bibitem{Genov2017} G. T. Genov, D. Schraft, N. V. Vitanov, and T. Halfmann, Arbitrarily Accurate Pulse Sequences for Robust Dynamical Decoupling, Phys. Rev. Lett. {\bf 118}, 133202 (2017).

\bibitem{Kolodrubetz2017} M. Kolodrubetz, D. Sels, P. Mehta, and A. Polkovnikov, Geometry and non-adiabatic response in quantum and classical systems, Phys. Rep. {\bf 697}, 1 (2017).

\bibitem{SM}Supplemental Material

\bibitem{Lindblad} G. Lindblad, On the generators of quantum dynamical semigroups, Commun. Math. Phys. $\textbf{48}$, 119-130 (1976).

\bibitem{Buluta2011} I. Buluta, S. Ashhab, and F. Nori, Natural and artificial atoms for quantum computation, Rep. Prog. Phys. {\bf 74}, 104401 (2011).

\bibitem{chen2016} Z. Chen, J. Kelly, C. Quintana, R. Barends, B. Campbell, Y. Chen, B. Chiaro, A. Dunsworth, A. G. Fowler, E. Lucero \emph{et al}., Measuring and Suppressing Quantum State Leakage in a Superconducting Qubit,
Phys. Rev. Lett. {\bf 116}, 020501 (2016).

\bibitem{Krantz2016} P. Krantz, M. Kjaergaard, F. Yan, T. P. Orlando, S. Gustavsson, and W. D. Oliver, A quantum engineer's guide to superconducting qubits, Appl. Phys. Rev. {\bf 6} 021318 (2019)


\bibitem{Song2016} X. K. Song, H. Zhang, Q. Ai, J. Qiu, and F. G. Deng, Shortcuts to adiabatic holonomic quantum computation in decoherence-free subspace with transitionless quantum 
driving algorithm, New J. Phys. {\bf 18}, 023001 (2016).

\bibitem{liang2014}Z.-T. Liang, Y.-X. Du, W. Huang, Z.-Y. Xue, and H. Yan, non-adiabatic holonomic quantum computation in decoherence-free subspaces with trapped ions, Phys. Rev. A {\bf 89}, 062312 (2014).

\bibitem{Feng2009} X.-L. Feng, C. Wu, H. Sun, and C. H. Oh, Geometric Entangling Gates in Decoherence-Free Subspaces with Minimal Requirements, Phys. Rev. Lett. {\bf 103}, 200501(2009).


\bibitem{Bravyi} S. B. Bravyi and A. Y. Kitaev, Quantum codes on a lattice with boundary, arXiv:quant-ph/9811052.

\bibitem{Dennis2002}  E. Dennis, A. Y. Kitaev, A. Landahl, and J. Preskill, Topological quantum memory, J. Math. Phys. {\bf 43}, 4452 (2002).

\bibitem{Fowler2012} A. G. Fowler, M. Mariantoni, J. M. Martinis, and A. N. Cleland, Surface codes: Towards practical large-scale quantum computation, Phys. Rev. A {\bf 86}, 032324 (2012).

\bibitem{ZhangJ2018}  J. Zhang, S. J. Devitt, J. Q. You, and F. Nori, Holonomic surface codes for fault-tolerant quantum computation Phys. Rev. A {\bf 97}, 022335 (2018). 

\bibitem{Wu2020}  C. Wu, Y. Wang, X.-L. Feng, and J.-L. Chen,  Holonomic Quantum Computation in Surface Codes, Phys. Rev. Applied {\bf 13}, 014055 (2020).

\bibitem{NC2018} K. Arai, J. Lee , C. Belthangady, D. R. Glenn, H. Zhang, and  R.L. Walsworth, Geometric phase magnetometry using a solid-state spin, Nat. Commun. {\bf 9}, 4996 (2018).




\end{thebibliography}
\end{document}